\documentclass[12pt,preprint]{aastex}
\usepackage{lscape}

\def\rel{{\rm rel}}
\def\e{{\rm E}}
 
\def\muas{{\mu\rm as}}

\def\kpc{{\rm kpc}}

\def\rel{{\rm rel}}

\def\mas{{\rm mas}}

\def\e{{\rm E}}

\newcommand{\grad}{\hbox{$^\circ$}}

\begin{document}
\title{Microlensing Event MOA-2007-BLG-400: Exhuming the Buried Signature of a Cool, Jovian-Mass Planet}

\author{
Subo~Dong\altaffilmark{1,2},
I.A.~Bond\altaffilmark{3,4},
A.~Gould\altaffilmark{1,2},
Szymon~Koz\l owski\altaffilmark{1,2},
N.~Miyake\altaffilmark{3,5},
B.S.~Gaudi\altaffilmark{2},
D.P.~Bennett\altaffilmark{3,6},\\ 
and\\
F.~Abe\altaffilmark{5}, 
A.C.~Gilmore\altaffilmark{7},
A.~Fukui\altaffilmark{5}, 
K.~Furusawa\altaffilmark{5},
J.B.~Hearnshaw\altaffilmark{7}, 
Y.~Itow\altaffilmark{5}, 
K.~Kamiya\altaffilmark{5}, 
P.M.~Kilmartin\altaffilmark{8}, 
A.~Korpela\altaffilmark{9}, 
W.~Lin\altaffilmark{7},
C.H.~Ling\altaffilmark{7},
K.~Masuda\altaffilmark{5}, 
Y.~Matsubara\altaffilmark{5}, 
Y.~Muraki\altaffilmark{10}, 
M.~Nagaya\altaffilmark{5}, 
K.~Ohnishi\altaffilmark{11}, 
T.~Okumura\altaffilmark{5}, 
Y.C.~Perrott\altaffilmark{12}, 
N.~Rattenbury\altaffilmark{13}, 
To.~Saito\altaffilmark{14}, 
T.~Sako\altaffilmark{5}, 
S.~Sato\altaffilmark{15}, 
L.~Skuljan\altaffilmark{4},
D.J.~Sullivan\altaffilmark{9}, 
T.~Sumi\altaffilmark{5}, 
W.~Sweatman\altaffilmark{7},
P.J.~Tristram\altaffilmark{8}, 
P.C.M.~Yock\altaffilmark{12},\\ 
({The MOA Collaboration})\\
G.~Bolt\altaffilmark{16},
G.W.~Christie\altaffilmark{17},
D.L.~DePoy\altaffilmark{2},
C.~Han\altaffilmark{18},
J.~Janczak\altaffilmark{2},
C.-U.~Lee\altaffilmark{19},
F.~Mallia\altaffilmark{20},
J.~McCormick\altaffilmark{21},
B.~Monard\altaffilmark{22},
A.~Maury\altaffilmark{20},
T.~Natusch\altaffilmark{23},
B.-G.~Park\altaffilmark{19},
R.W.~Pogge\altaffilmark{2},
R.~Santallo\altaffilmark{24},
K.Z.~Stanek\altaffilmark{2}\\
(The $\mu$FUN Collaboration),\\
A.~Udalski\altaffilmark{25},
M.~Kubiak\altaffilmark{25},
M.K.~Szyma{\'n}ski\altaffilmark{25},
G.~Pietrzy{\'n}ski\altaffilmark{25,26},
I.~Soszy{\'n}ski\altaffilmark{25},
O.~Szewczyk\altaffilmark{25,26},
{\L}.~Wyrzykowski\altaffilmark{25,27},
{K}.~Ulaczyk\altaffilmark{25}\\
(The OGLE Collaboration)\\
}
\altaffiltext{1}
{Microlensing Follow Up Network ($\mu$FUN)}
\altaffiltext{2}
{Department of Astronomy, Ohio State University,
140 W.\ 18th Ave., Columbus, OH 43210, USA; 
dong,gould,simkoz,gaudi,depoy,pogge,kstanek@astronomy.ohio-state.edu}
\altaffiltext{3}
{Microlensing Observations for Astrophysics (MOA) Collaboration}
\altaffiltext{4}
{Institute of Information and Mathematical Sciences, Massey University,
Private Bag 102-904, North Shore Mail Centre, Auckland, New Zealand;
i.a.bond@massey.ac.nz}
\altaffiltext{5}
{Solar-Terrestrial Environment Laboratory, Nagoya University, 
Nagoya, 464-8601, Japan}
\altaffiltext{6}
{Department of Physics, Notre Dame University, Notre Dame, IN 46556, USA;
bennett@nd.edu}
\altaffiltext{7}
{University of Canterbury, Department of Physics and Astronomy, 
Private Bag 4800, Christchurch 8020, New Zealand.}
\altaffiltext{8}{Mt. John Observatory, P.O. Box 56, Lake Tekapo 8770, New Zealand.}
\altaffiltext{9}{School of Chemical \& Physical Sciences, Victoria University, Wellington, P.O. Box 600 Wellington 6012, New Zealand; denis.sullivan@vuw.ac.nz}
\altaffiltext{10}{Department of Physics, Konan University, Nishiokamoto 8-9-1, 
Kobe 658-8501, Japan.}
\altaffiltext{11}{Nagano National College of Technology, Nagano 381-8550, Japan.}
\altaffiltext{12}
{Department of Physics, University of Auckland, 
Private Bag 92019, Auckland, New Zealand;
yper006@ec.auckland.ac.nz,p.yock@auckland.ac.nz}
\altaffiltext{13}
{Jodrell Bank Centre for Astrophysics, The University of Manchester,
United Kingdom;
njr@jb.man.ac.uk}
\altaffiltext{14} 
{Tokyo Metropolitan College of Industrial Technology, Tokyo 116-0003, Japan}
\altaffiltext{15}
{Department of Physics and Astrophysics, Faculty of Science,
Nagoya University, Nagoya 464-8602, Japan}
\altaffiltext{16} 
{Center for Backyard Astrophysics, Perth, Australia, gbolt@iinet.net.au}
\altaffiltext{17}
{Auckland Observatory, Auckland, New Zealand, gwchristie@christie.org.nz}
\altaffiltext{18}
{Department of Physics, Institute for Basic Science Research,
Chungbuk National University, Chongju 361-763, Korea;
cheongho@astroph.chungbuk.ac.kr}
\altaffiltext{19}
{Korea Astronomy and
Space Science Institute, Daejon 305-348, Korea; leecu,bgpark@kasi.re.kr}
\altaffiltext{20}
{Campo Catino Austral Observatory, San Pedro de Atacama, Chile}
\altaffiltext{21}
{Farm Cove Observatory, Centre for Backyard Astrophysics,
Pakuranga, Auckland New Zealand; farmcoveobs@xtra.co.nz}
\altaffiltext{22} 
{Bronberg Observatory, South Africa, lagmonar@nmisa.org}
\altaffiltext{23} 
{Institute for Radiophysics and Space Research, AUT University, Auckland,
New Zealand, tim.natusch@aut.ac.nz}
\altaffiltext{24} 
{Southern Stars Observatory, Tahiti, obs930@southernstars-observatory.org}
\altaffiltext{25}
{Warsaw University Observatory, Al.~Ujazdowskie~4, 00-478~Warszawa,Poland; 
udalski,soszynsk,msz,mk,pietrzyn,szewczyk,wyrzykow,kulaczyk@astrouw.edu.pl}
\altaffiltext{26}{Universidad de Concepci{\'o}n, Departamento de Fisica,
Casilla 160--C, Concepci{\'o}n, Chile}
\altaffiltext{27} {Institute of Astronomy  Cambridge University,
Madingley Rd., CB3 0HA Cambridge, UK wyrzykow@ast.cam.ac.uk}

\begin{abstract}
We report the detection of the cool, Jovian-mass planet MOA-2007-BLG-400Lb.
The planet was detected in a high-magnification microlensing event (with
peak magnification $A_{\rm max} = 628$) in
which the primary lens transited the source, resulting in a dramatic
smoothing of the peak of the event.  The angular extent of the region
of perturbation due to the planet is significantly smaller than the
angular size of the source, and as a result the planetary signature
is also smoothed out by the finite source size.  Thus the deviation
from a single-lens fit is broad and relatively weak ($\sim$ few
percent).  Nevertheless, we demonstrate that the planetary nature 
of the deviation can be
unambiguously ascertained from the gross features of the residuals,
and detailed analysis yields a fairly precise planet/star mass ratio of
$q=2.6\pm 0.4 \times 10^{-3}$, in accord with the large significance
($\Delta\chi^2=1070$) of the detection.  The planet/star projected
separation is subject to a strong close/wide degeneracy, leading to
two indistinguishable solutions that differ in separation by a
factor of $\sim 8.5$.  Upper limits on flux from the lens constrain
its mass to be $M<0.75\,M_\odot$ (assuming it is a main-sequence star). 
A Bayesian analysis that includes all available
observational constraints indicates a primary in the Galactic bulge
with a mass of $\sim 0.2-0.5 M_\odot$ and thus a planet mass of $\sim
0.5-1.3 M_{\rm Jup}$.  The separation and equilibrium temperature are
$\sim 0.6-1.1~{\rm AU}$ ($\sim 5.3-9.7~{\rm AU}$) and $\sim 103~{\rm K}$
($\sim 34~{\rm K}$) for the close (wide) solution. If the primary is
a main-sequence star, follow-up observations would enable the
detection of its light and so a measurement of its mass
and distance.
\end{abstract}

\section{Introduction
\label{sec:intro}}

In the currently favored paradigm of planet formation, the location of
the snow line in the protoplanetary disk plays a pivotal role.  Beyond
the snow line, ices can condense, and the surface density of solids is
expected to be higher by a factor of several relative to its value
just inside this line.  As a result of this increased surface density,
planet formation is expected to be most efficient just beyond the snow
line, whereas for increasing distances from the central star the
planet formation efficiency drops, as the surface density decreases
and the dynamical time increases \citep{lissauer87}.  In this
scenario, gas-giant planets must form in the region of the
protoplanetary disk immediately beyond the snow-line, as the higher
surface density is required to build icy protoplanetary cores that
are sufficiently massive to accrete a substantial gaseous envelope
while there is remaining nebular gas \citep{pollack96}.  Low-mass
primaries are expected to be much less efficient at forming gas giants
because of the longer dynamical times and lower surface densities at
the snow lines of these stars \citep{laughlin04,ida05,kennedy08}.
Migration due to nebular tides and other dynamical processes can then
bring the icy cores or gas giants from their formation sites to orbits
substantially interior to the snow line \citep{lin96,ward97,rasio96}.

The precise location of the snow line in protoplanetary disks is a
matter of some debate (e.g., \citealt{lecar06}), and is even likely to
evolve during the epoch of planet formation, particularly for
low-mass stars \citep{kennedy06,kennedy08}.  The condensation
temperature of water is $\sim 170~{\rm K}$, and a fiducial value for
the location of the snow line in solar-mass stars motivated by
observations in our solar system is $\sim 2.7~{\rm AU}$.  This may
scale linearly with the stellar mass $M$, since the stellar luminosity
during the epoch of planet formation scales as $\sim M^2$ for stars with 
$M \la M_\odot$ \citep{burrows93,burrows97}.  Whereas the radial velocity and
especially transit methods are most sensitive to planets that are
close to their parent star at distances well inside the snow line, the
sensitivity of the microlensing method peaks at planetary separations
near the Einstein ring radius of the primary lens 
\citep{mp91,gould92}, which is $\sim 3.5~{\rm AU}(M/M_\odot)^{1/2}$
for typical lens and source distances of $6~{\rm kpc}$ and $8~{\rm
kpc}$, respectively.  This corresponds to a peak sensitivity at
equilibrium temperatures of $T_{\rm eq} \sim 150~{\rm K} (M/M_\odot)$
for a mass-luminosity relation of the form $L \propto M^{5}$, and
distances relative to the snow line of $\sim 1.3 (M/M_\odot)^{-1/2}$ 
if the location of the snow line at the epoch of planet formation
scales as $M$.
Thus microlensing is currently the best method of probing planetary
systems in the critical region just beyond the snow line
\citep{gould92}.

Planetary perturbations in microlensing events come in two general
classes.  The majority of planetary perturbations are expected to
occur when a planet directly perturbs one of the two images created by
the primary lens, as the image sweeps by the planet during the
microlensing event \citep{gould92}.  Although these perturbations are
more common, they are also unpredictable and can occur at any time
during the event.  Early microlensing planet searches focused on this
class of perturbations, as it was the first to be identified and
explored theoretically \citep{gould92,bennett96,gaudi97}.  The second
class of planetary perturbations occurs in high-magnification events,
in which the source becomes very closely aligned with the primary lens
\citep{griestsafi}.  In such events, the two primary-lens images
become highly distorted and sweep along nearly the entirety of the
Einstein ring \citep{liebes}.  These sweeping images
probe subtle distortions of the Einstein ring caused by nearby
planets, which will give rise to perturbations within the full-width
half-maximum of the event \citep{bond02,rattenbury02}.  Although
high-magnification events are rare and so contribute a minority of
the planetary perturbations, they are individually more sensitive
to planets because the images probe nearly the entire Einstein ring.
Furthermore, since the perturbations are localized to the peak of the
event which can be predicted beforehand, they can be monitored more
efficiently with limited resources than the more common
low-magnification events.

For these reasons, current microlensing planet searches tend to
deliberately focus on high-magnification events.  Thus, of the seven
prior microlensing planets discovered to date
\citep{ob03235,ob05071,ob05390,ob05169,ob06109,mb07192}, five have
been found in high-magnification events, with peak magnifications
ranging from $A=40$ to $A=800$.  However, despite the fact that this
planet-search strategy has proven to be so successful, the properties
of the planetary perturbations generated in high-magnification events
are less well-understood than those in low-magnification events.

Most of the studies of the properties of planetary perturbations in
high-magnification events have focused on the properties of the
caustics, the locus of points defining one or more closed curves, upon
which the magnification of a point source is formally infinite.  The
morphology and extent of the region of significant perturbation by the
planetary companion can be largely understood by the shape and size of
these caustic curves.  Planetary perturbations in high-magnification
events are caused by a central caustic located near the
position of the primary, and thus several authors have considered the
size and shape of these central caustics as a function of the
parameters of the planet \citep{griestsafi,dominik99,chung05}.  
These and other authors have identified several potential degeneracies
that complicate the unique interpretation of central caustic
perturbations.  The first to be identified is a degeneracy such that
the caustic structure (and so light curve morphology) of a planet with
mass ratio $q \ll 1$ and projected separation in units of the
Einstein ring $d$ not too close to unity is essentially identical under
the transformation $d \leftrightarrow d^{-1}$ \citep{griestsafi}.  A
second degeneracy arises from the fact that very close or very wide
roughly equal-mass binaries also produce perturbations near the peak
of the light curves.  These give rise to perturbations
that have the same gross observables as planetary perturbations.

The severity of these degeneracies depends on both the specific
parameters of the planetary/binary companion, as well as on the data
quality and coverage. \citet{griestsafi} and \cite{chung05}
demonstrated that the $d \leftrightarrow d^{-1}$ degeneracy is less
severe for more massive planets with separations closer to the
Einstein ring ($d \sim 1$).  Empirically, this degeneracy was broken
at the $\Delta \chi^2 \sim 4$ level for the relatively large
mass-ratio planetary companion OGLE-2005-BLG-071Lb \citep{dong08}, for
which the light curve was well-sampled, but was essentially unresolved
for the low mass-ratio planetary companion MOA-2007-BLG-192Lb
\citep{mb07192}, for which the planetary perturbation was poorly
sampled.  For the planetary/equal-mass binary degeneracy,
\citet{han08} argued that, although the gross features of central
caustic planetary perturbations can be reproduced by very close or
very wide binary lenses, the morphologies differ in detail, and thus
this degeneracy can be resolved with reasonable light curve coverage
and moderate photometric precision. Indeed for every well-sampled
high-magnification event containing a perturbation 
near the peak (and that is not in the \citet{cr1} limits), this
degeneracy has been resolved \citep{albrow02,ob05169,dong08}.  Even for
the relatively poorly-sampled light curve of MOA-2007-BLG-192Lb, an
equal-mass binary model is ruled out at $\Delta\chi^2\sim 120$
\citep{mb07192}, although in this case this is partially attributable
to the exquisite photometric precision ($<1\%$).
 
One complication with searching for planets in high-magnification
events is that, the higher the magnification, the more likely it is
that the primary lens will transit the source.  When this happens, the
peak of the event is suppressed and smoothed out, as the lens strongly
magnifies only a small portion of the source. If the source is also
larger than the region of significant perturbations due to a planetary
companion (roughly the size of the central caustic), then the
planetary deviations will also be smoothed out and suppressed
\citep{griestsafi,han07}. These finite source effects have potential
implications for both the detectability of central caustic
perturbations, as well as the ability to uniquely determine the
planetary parameters, and in particular resolve the two degeneracies
discussed above.  In practice, the caustic structures of all four
high-magnification planetary events (containing five planets) were
larger than the source.  Hence, while there were detectable
finite-source effects in all cases (which helped constrain the angular
Einstein radius and so the physical lens parameters), the planetary
perturbations were in all four cases quite noticeable.  Thus the
effect of large sources on the the detectability and interpretation of
central-caustic perturbations has not been explored in practice.

Theoretical studies of detectability of central caustic perturbations
when considering finite source effects have been performed by
\citet{griestsafi}, \citet{chung05}, and \citet{han07}.  These authors
demonstrated that the qualitative nature of planetary perturbations
from central caustics is dramatically different for sources that are
larger than the caustic.  In particular, the detailed structure of the
point-source magnification pattern, which generally follows the shape
of the caustic, is essentially erased or washed out.  Rather, the
perturbation structure is characterized by a roughly circular region
of very low-level, almost imperceptible deviation from the single-lens form that is
roughly the size of the source and centered on the primary lens.  This
region is surrounded by an annular rim of larger deviations that has
a width roughly equal to the width of the caustic
\citep{griestsafi,chung05}. Finally, there are less pronounced deviations
that extend to a few source radii. Planets are detectable even if their
central caustics are quite a bit smaller than the source, provided
that the $\chi^2$ deviation is sufficiently high.  (Often
$\Delta\chi^2>60$ is adopted, although $\Delta\chi^2>150$ may be more
realistic.)  The magnitude of these perturbations decrease as the
ratio between the source size and caustic size increases, making it
difficult to detect very small planets for large sources
\citep{chung05,han07}.

Although central caustics may formally be detectable when the source
is substantially larger than the caustic, it remains a significant
question whether these very washed out caustics can be
recognized in practice, and even if they can, whether they can be
uniquely interpreted in terms of planetary parameters.  Indeed, it is
unknown whether a washed out central caustic due to a planet can
actually be distinguished from one due to a binary companion.  This
question is especially important with regard to low-mass planets.  The
size of the central caustic scales as the product of the planet/star
mass ratio and a definite function of planet-star separation.  Hence,
taken as a whole, smaller planets produce smaller caustics, meaning
that events of higher magnification are required to detect them.
These are just the events that are most likely to have their peaks
washed out by finite-source effects.

Here we analyze the first high-magnification event with a buried
signature of a planet, in which the source size is larger than the
central caustic of the planet.  The caustic is indeed so washed out
that the event appears unperturbed upon casual inspection.  However,
the residuals to a point-lens fit are clear and highly significant.
We show that one can infer the planetary (as opposed to binary) nature
of the perturbation from the general pattern of these residuals, and
that a detailed analysis constrains the mass ratio of the planet quite
well, but leaves the close/wide ($d \leftrightarrow d^{-1}$) degeneracy intact.
Hence, at least in this case, the fact that the caustic is buried in
the source does not significantly hinder one's ability 
to uncover the planet and measure its mass ratio.

\section{Observations
\label{sec:obs}}

MOA-2007-BLG-400 [$(\alpha, \delta)_{{\rm J}2000.0}=(18^{\rm h}09^{\rm m}41.\!\!^{\rm s}98,-29\grad 13^\prime 26.95^{\prime\prime})$,
(l,b)=(2.38,$-4.70$)]
was announced as a probable microlensing event by
the Microlensing Observations in Astrophysics (MOA) collaboration
on 5 Sept 2007 (HJD$' \equiv$ HJD - 2450000 = 4349.1), about 5 days
before peak.  The source star proves to be a bulge subgiant and so
is somewhat brighter than average, but the event timescale was
relatively short ($t_\e\sim 15\,$days) and observations had been
interrupted for 6 days by bad weather.  Taken together, the two latter
facts account for
the relatively late alert.  By coincidence, the triggering observations
took place on the same night that another event, 
OGLE-2007-BLG-349 (aka MOA-2007-BLG-379),
was peaking at extremely high magnification with an already-obvious
planetary anomaly.  After focusing exclusively on the latter event for
the first 5 hours of the night, MOA resumed its normal field rotation
for the last 1.5 hours, which led to the discovery of MOA-2007-BLG-400.

The Microlensing Follow Up Network ($\mu$FUN) began observing this as a
possible high-magnification event on 7 Sept, but did not mobilize intensive
observations until UT 08:55 10 Sept, just 15 hours before peak,
following a high-mag alert issued by MOA a few minutes earlier.  Even
at that point, the predicted minimum peak magnification was only 
$A_{\rm max}>90$, which would have enabled only modest sensitivity
to planets.  Nevertheless, all stops were pulled and it was observed
as intensively as possible from 7 observatories,
$\mu$FUN CBA Perth (Australia) 0.25m unfiltered,
$\mu$FUN Bronberg (South Africa) 0.35m unfiltered, 
$\mu$FUN SMARTS (CTIO, Chile) 1.3m $V$, $I$, $H$,
$\mu$FUN Campo Catino Austral (CAO, Chile) 0.50m unfiltered,
$\mu$FUN Farm Cove (New Zealand) 0.35m unfiltered,
$\mu$FUN Auckland (New Zealand) 0.4m $R$, and
$\mu$FUN Southern Stars (Tahiti) 0.28m unfiltered.

The source star lies just outside one of the OGLE fields (as defined
by their field templates) and for this reason was not recognized as
a microlensing event
by the OGLE Early Warning System.  However, due to small variations
in pointing, there are a total of 452 OGLE images containing this source.
Only two of these are {\it significantly} magnified, ten days and nine days 
before peak.
Hence, the OGLE data do not help constrain the light curve parameters.
However, they are useful to study of the baseline behavior of the
source (see Appendix~\ref{sec:append_ell}).

Essentially all of the ``action'', both the peak of the
event and the planetary anomalies, occurred during the $\mu$FUN
SMARTS (Chile) observations at CTIO, $4354.47 < {\rm HJD}' < 4354.69$, using
the ANDICAM optical/IR dual-channel camera, and $\mu$FUN CAO (Chile)
observations $4354.50 < {\rm HJD}' < 4354.70$.
Most (45) of the optical CTIO observations over the peak
were carried out in $I$ band, with a few (8) taken in $V$ in order
to measure the $(V-I)$ color.  Each of these was a 5 minute
exposure, with approximately 1 minute read-out time between exposures.  
During each optical exposure, there were 5 dithered
$H$-band exposures,
each of 50 seconds, almost equally spaced over the 6 minute cycle
time.  That is, $53\times 5=265$ $H$-band observations in all.  
{\rm Unfortunately, the source became so bright as it transited
the lens (i.e., when the planetary anomalies were the strongest),
that 14 $I$-band images were affected by non-linearities and saturation
in the detector response. We exclude these 14 $I$-band data points
from analysis. The $H$-band photometry are not affected by this problem, 
therefore, with higher time resolution and more continuous
coverage than the $I$-band data, the $H$-band data provide most of the
constraining power to the microlens model.} 

There were 84 CAO un-filtered observations taken during the peak night. 
Unfortunately, the peak of the event was severely saturated and
the clock zero point is not securely known, therefore, these data are not
used in the analysis.  However, the
exposures during the times of maximum deviation from a point lens
(i.e., when the caustic was crossing the stellar limb) are not
saturated, and these qualitatively confirm the interpretation from the more
detailed CTIO data.

The MOA data were reduced using the standard MOA difference imaging
analysis (DIA) pipeline. All $\mu$FUN data were reduced using DIA 
developed by \citet{wozniak}. 
The $H$-band data are affected by intrapixel sensitivity variations
at the 1\% level.  Fortunately, the dither pattern was repeated
almost exactly over the night of the peak, 
so that these variations follow the 5-element
dither pattern quite well.  We therefore treat the $H$-band
data as 5 independent data sets, which reduces $\chi^2$ by 180
for 8 degrees of freedom. The $H$-band images show a triangular
PSF, which is likely to introduce systematic errors into the photometry. 
As a crosscheck, we also use the DIA package developed by \citet{bond01} to 
independently reduce these images.

\section{Microlens Model
\label{sec:model}}

Despite the fact that the peak of MOA-2007-BLG-400 was ``flattened''
by finite-source effects, it nevertheless reached a very high
peak magnification, $A_{\rm max}=628$.
However, even to the experienced eye, it
looks like an ordinary point-lens light curve with pronounced
finite-source effects.  
More detailed modeling is required
to infer that it actually contains a Jovian mass-ratio
planet.

\begin{figure}
\plotone{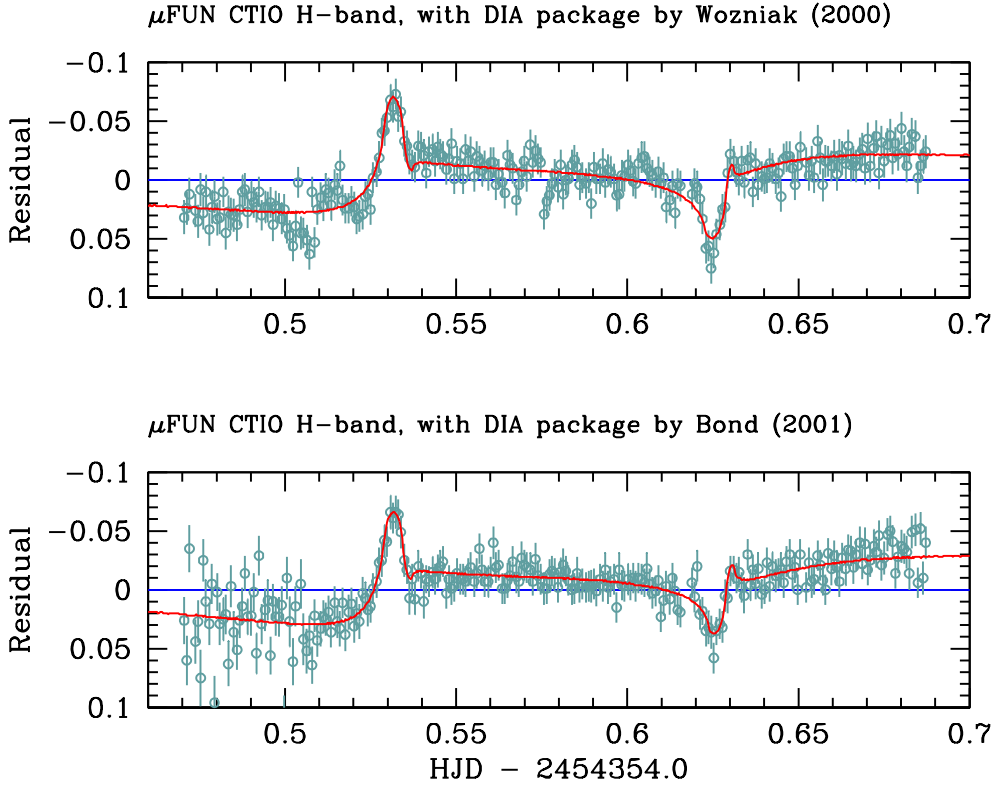}
\caption{Comparisons of residuals to the best-fit point-lens models 
between two photometric reductions of $\mu$FUN $H$-band data, using the
DIA packages developed by \citet{wozniak} (top) and \citet{bond01} (bottom).
The red curve in each panel represents the difference 
of the best-fit planetary and point-lens models for that panel's reduction. 
Both reductions agree on the main planetary features, but each
package introduces its own systematics. For example, the systematic 
deviations from the planetary model at ${\rm HJD} \sim 2454354.57$ shown in the 
top panel are not supported by the reductions of the Bond's package. During 
$2454354.46 <{\rm HJD }< 2454354.51$, most images have low transparency 
($\la 50\%$), which causes relatively large scatter in Bond's DIA reductions.
In comparison, the reduction by Wozniak's DIA has smaller scatter  
during this period. However, the data exhibit some low-level systematics, which
are not supported by the other reduction.}
\label{fig:res}
\end{figure}

Figure \ref{fig:lc} shows the light curve together with the best-fit
point-lens model (blue) and planetary model (red).  Both include
finite-source effects.  The most pronounced features of the point-lens
model residuals are a short positive spike on the rising side and a
short negative spike on the falling side, each lasting about 30
minutes, which leave very similar traces in $I$ and $H$. 
As displayed in Figure \ref{fig:res}, these features clearly stand out
in the reductions using both the Wozniak (top panel) and Bond (bottom panel) 
DIA packages. Each package introduces its own systematic deviations, but 
there are no obvious trends besides the above features that are 
supported by both reductions.
These occur very close to the times that the point lens
begins and ends its transit of the source (within the framework of this
model).  The timing of these deviations strongly suggests that 
they are due to microlensing rather than stellar variability. There are then
two possible explanations: either the source is actually being transited 
by a more complicated caustic than a point lens (due to a binary or planetary
companion) or the limb of the source is not being properly modeled.
However, if one assumes a circular source, 
the latter explanation would imply symmetric residuals,
whereas the actual residuals are closer to being antisymmetric.
(We address the possibility of an elliptical source in 
Appendix~\ref{sec:append_ell}.)\ \ 
Indeed, this approximate antisymmetry extends to the less pronounced
residual features, including the sustained deficit prior to the first
spike (and sustained excess following the second one) as well as the
declining residuals between the two spikes. The similarity of the
$I$ and $H$ residuals in itself argues that the deviations are
due to microlensing rather than some sort of stellar variability, 
which would not generally be expected to be achromatic. 

The short durations of the spikes tell us that the central caustic is
quite small, with ``caustic width'' $w\sim 30\,{\rm min}/15\,{\rm day} = 
1.4\times 10^{-3}$. This implies that the companion either has 
low mass ratio, or is a very wide or very close binary companion.
Formally, $w$ is given by equation (12) of \citet{chung05} as
a very good approximation to
the ``short diameter'' or ``width'' of a central
caustic (see Fig.~\ref{fig:caustic}).  However, 
here the estimate is quite inexact not only because
the width of the spikes is not precisely defined, but also because we do not
know, at this point, the exact orientation of the caustic.

After some algebra, one finds that in two limiting regimes, the 
\citet{chung05} formula takes the forms,
\begin{equation}
w(d,q) \rightarrow {4q\over d^2} \quad (d\gg 1),\qquad
w(d,q) \rightarrow {4q d^2} \quad (d\ll 1),
\label{eqn:bgg1}
\end{equation}
and
\begin{equation}
w(d,q) \rightarrow \sqrt{27\over 16}\,{q\over |d-1|} 
\qquad (d\sim 1),
\label{eqn:bsim1}
\end{equation}
where $q$ is the
companion/primary mass ratio and  $d$ is the separation in units
of the Einstein ring.  Note that in the first limit (eq.~[\ref{eqn:bgg1}]),
$w\rightarrow 4\gamma$, where $\gamma$ is the shear.
The crossover point for these approximations
is $d=2.3$ (or $d=0.43$), at which point each is in error by about 15\%.
(For simplicity, we restrict the discussion here to
the case $d>1$.  There is a well-known $d\leftrightarrow d^{-1}$ degeneracy
between the $d \ll 1$ and $d \gg 1$ limits, as can be guessed from the 
forms for w in 
eq.~[\ref{eqn:bgg1}]. This degeneracy will prove to be almost perfect 
in this case, see \S~\ref{sec:broadsearch}).\ \ 
Hence, in the first limit, $d\sim 50 q^{1/2}$, implying that if
$q$ were in the ``binary range'' ($|\log q|<1$), then $d$ would be quite large.
That is, the central caustic would be generated by a nearly pure shear
and therefore would have a nearly symmetric, diamond-shaped, \citet{cr1}
form.  In the point-lens model, the lens passes almost directly
over the center of the source.  For this trajectory, a symmetric
caustic would yield symmetric residuals, in sharp contrast to
Figure \ref{fig:lc}.  On the other hand, for the opposite limit,
$q\sim 2\times 10^{-3}(d-1)$, which lies squarely in the planetary regime.
Thus, simple arguments already argue strongly in favor 
of a planetary companion that is fairly near the Einstein ring.

\subsection{Hybrid Pixel/Ray Map Algorithm
\label{sec:algorithm}}

Notwithstanding these arguments, we conduct a massive blind search
for companions over a very broad range of masses using a 
modified version of the  ``magnification map'' technique of \citet{ob04343},
which was specifically designed for high-magnification events.
The original approach was, for each given $(d,q)$ pair,
to shoot rays over a fairly narrow annulus
(say, 0.01 Einstein radii) around the Einstein ring in the image
plane and to sort these rays in pixels on the source plane.  Then for
each source position being modelled (i.e., each data point), one
would identify the pixels that intersected the source and
would check each ray contained in these pixels to determine whether
it landed on the source and, if so, evaluate the source surface brightness
at that position.  In the initial broad search, three parameters
$(d,q,\alpha)$ are held fixed on a grid of values, while the remaining
parameters ($t_0,u_0,t_\e,\rho$, and possibly others) 
are varied to minimize $\chi^2$ at each grid 
point.  Here, $\alpha$ is the angle of the source trajectory relative to the 
binary axis, $t_0$ is the time of closest approach to the adopted
center of the lens geometry (usually the center of mass), $u_0$ is the
source-lens separation at this time in units of the Einstein radius,
$\rho$ is the source radius in the same units, and $t_\e$ is the
Einstein crossing time.  This division is efficient because
1) ($t_0,u_0,t_\e,\rho$) are usually approximately known from the
general structure of the light curve, so minimization over these
parameters is straightforward once $(d,q,\alpha)$ are fixed; 
2) $(d,q)$ define the map, which naturally facilitates minimization
of other parameters except for $\alpha$, whose value is not usually 
even approximately obvious from the light curve.

The new approach differs principally in that the pixels that are contained
entirely within the source are now evaluated as a whole, i.e., by
the total number of rays in that pixel.  Pixels that cross the source
boundary are still evaluated ray-by-ray, as previously.  This primary
change then leads to several other changes.  First, for each pixel,
we record not only the geometric center but also the centroid 
of the rays.  The surface brightness is then evaluated at the latter
position.  Second, the pixels are made much smaller, to minimize both
the number of rays that must be evaluated individually and 
the surface-brightness variations across the pixel (which are corrected 
only to first order by
the ray-centroid scheme just mentioned).  Typically, there are a few hundreds of pixels per source.  Third, the pixels are hexagonal, since this is
the most compact tiling possible, i.e., the closest tile shape to a
circle.  Fourth, the source positions outside the map region are
evaluated using the hexadecapole and quadrupole approximation of 
\citet{gould08} (see also \citealt{pejcha07}). 
Finally, we use Markov chain Monte Carlo (MCMC) for 
the $\chi^2$ minimization.

\subsection{The $(w,q)$ Grid of Lens Geometries
\label{sec:wq}}

The initial search for solutions is conducted over a rectilinear
grid in $(w,q)$ rather $(d,q)$.  Since the short diameter,
or ``caustic width'', $w$, is a monotonic function of the star-planet
separation $d$ (at fixed planet/star mass ratio $q$), these 
formulations are in some sense equivalent.  However, for many events
(including the present one) the short diameter $w$ can be estimated
by simple examination of the data.  In these cases, the search space
is both more regular and easier to define in terms of the $(w,q)$ grid.
In particular, equation (\ref{eqn:bsim1}) shows that at fixed $w$,
$d$ moves very close to 1 for very low $q$.

\subsection{Best-Fit Model
\label{sec:broadsearch}}

We consider
short diameters $w$ over the range 
$-3.5\leq\log w\leq -2$ and companion mass ratios
$-4\leq\log q\leq 0$, focusing on the regime $d\ge 1$.  We find that there is only one
local minimum in this range.  The range of allowed solutions is well
localized around this minimum, with 
\begin{equation}
q = 2.6\pm 0.4\times 10^{-3},\qquad
d = 2.9\pm 0.2,\qquad
w = 1.30\pm 0.06\times 10^{-3},
\label{eqn:bqg}
\end{equation}
with the last quantity being, of course, dependent on the first two.
Figure \ref{fig:qgamma} shows the $\Delta\chi^2 = 1,\,4,\,9$ contours
with respect to the mass ratio $q$ and the projected planet-star separation
$d$ along with the short diameter $w$.
The main point to note is that these parameters are
quite well constrained. Note that, as expected, $d$ and $q$ are strongly correlated, while $w$ and $q$ are basically uncorrelated. 
We also perform similar searches using the alternative 
$H$-band reduction by Bond's DIA package. The solutions agree with
the above to well within one sigma, but the parameters have larger 
uncertainties: 
$q = 2.6\pm 0.7\times 10^{-3}, d = 2.9\pm 0.3, w = 1.26\pm 0.09\times 10^{-3}$.
We therefore adopt results from the Wozniak-based reductions, noting that they
may subject to systematic errors $\la 1 \sigma$.

Figure \ref{fig:caustic} shows the source trajectory and the central
caustics as well as the differences in magnification between the best-fit
planetary model and its corresponding single-lens model. This geometry
nicely accounts for the main features of the point-lens residuals
seen in Figure \ref{fig:lc}.  The regions beyond the ``back walls''
(long segments)
of the caustic are somewhat de-magnified, which accounts for the initial
depression of the light curve.  
As the source crosses the ``back wall'' of  the caustic,
it spikes.  After the source has exited the caustic, it continues to
suffer additional magnification due to the ``ridge'' of magnification
that extends from the trailing cusp.  

We also conducted a similar blind search as above, but concentrating
on the regime $d<1$. As expected, we recover the well-known
$d\leftrightarrow d^{-1}$ degeneracy, and find a solution with essentially the
same $q$, but with $d=2.9^{-1}=0.34 \pm 0.02$, and 
 the wide solution is slightly preferred
by $\Delta \chi^2 =0.2$.  Thus, although each solution is well-localized to its respective
minimum, this discrete degeneracy implies that the projected
separation can take on two values that differ by a factor of $\sim 8.5$.
The severity of the degeneracy can be traced to the
planetary parameters.  Although the planet/star mass ratio is quite
large, which tends to reduce the severity of the degeneracy, the
planet lies quite far from the Einstein ring, which tends to make it
more severe.  Actually, a better measure of the overall expected
asymmetry between the $d$ and $d^{-1}$ solutions is the short diameter $w$,
which in this case is small, implying a severe degeneracy.  Indeed,
the caustic structure and magnification pattern of the two solutions
are nearly identical. In this case, the large size of the source has 
competing influences on the ability to resolve the degeneracy.  On one
hand, the large size of the source serves to suppress the planetary
deviations, thus making subtle differences more difficult to
distinguish.  On the other hand, the large source implies that a large
fraction of the planetary perturbation region is probed.  In this
case, the source probes essentially the entire region of significant
planetary perturbation, as can be seen in Figure \ref{fig:caustic}.
This is important for distinguishing between the solutions, as the
largest difference between the magnification patterns of the two
degenerate solutions occurs in the region near the tip of the
arrow-shaped caustic \citep{griestsafi}.  From Figure \ref{fig:caustic} 
it is clear that
this region would have been entirely missed if the source had been
substantially smaller than the caustic.

\section{Finite-Source Effects
\label{sec:finitesource}}

\begin{figure}
\plotone{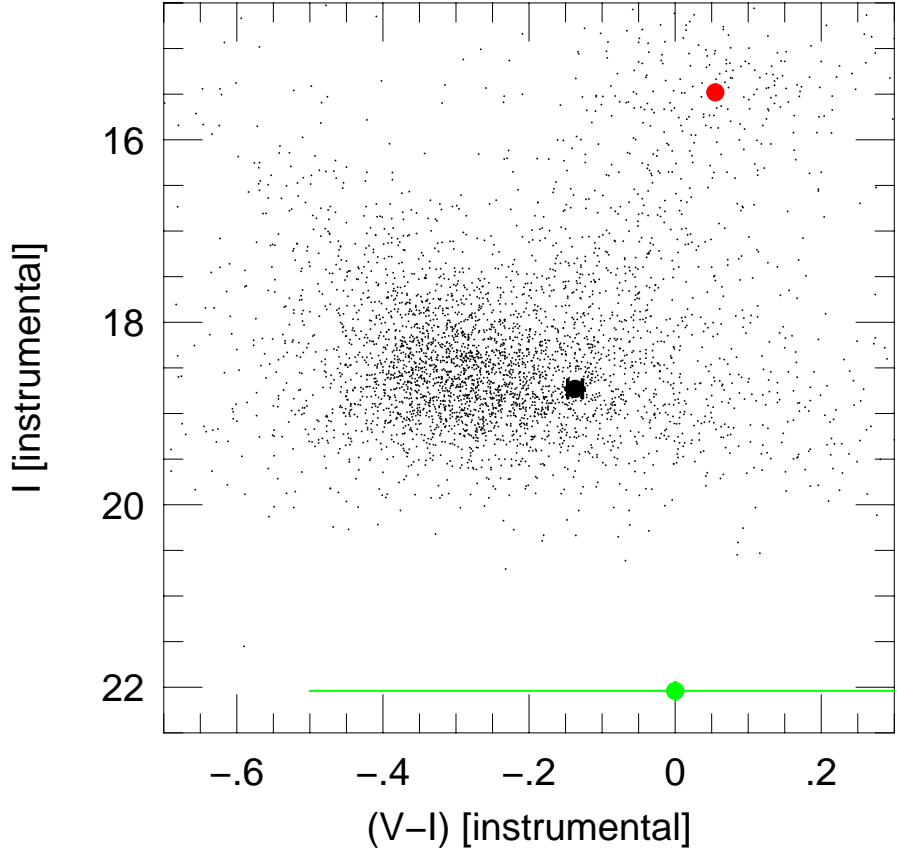}
\caption{Instrumental color-magnitude diagram of field containing
MOA-2007-BLG-400.  The color and magnitude of the source ({\it black})
are derived from the fit to the light curve, which also yields 
an {\it upper limit} for the
$I$-band blended flux ({\it green}).  The large error bar on the latter
point indicates a complete lack of information about its $V$-band flux.
The clump centroid is indicated in {\it red}.  From the source-clump
offset, we estimate $[I,(V-I)]_{0,s} = (17.57,0.81)$, implying it has
angular radius $\theta_*=1.05\,\muas$.
Assuming the source lies at 8 kpc, it has $[M_I,(V-I)]_{0,s} = (3.07,0.81)$,
making it a subgiant. The lack of blended light allows us to place 
an upper limit on the lens flux, which implies that it has 
mass $M<0.75\,M_\odot$.  See text.
}
\label{fig:cmd}
\end{figure}

In addition to $(d,q)$, the model also yields the source radius
relative to the Einstein radius, 
\begin{equation}
\rho={\theta_*/\theta_\e} = 3.29\pm 0.08\times 10^{-3},
\label{eqn:rho}
\end{equation}
We then follow the standard \citep{ob03262} technique to determine 
the angular source radius,
\begin{equation}
\theta_* = 1.05\pm 0.05\,\muas.
\label{eqn:thetastar}
\end{equation}
That is, we first adopt $[(V-I)_0,I_0]_{\rm clump} = (1.00,14.32)$ for
the dereddened position of the clump.  We then
measure the offset of the source relative to the clump
centroid $\Delta[(V-I),I] = (-0.19,3.25)$, to obtain 
$[(V-I)_0,I_0]_s = (0.81,17.57)$. See Figure \ref{fig:cmd}. 
The instrumental source color is derived from model-independent 
regression of the $V$ and $I$ flux,
while the instrumental magnitude is obtained from the light-curve model.
We convert $(V-I)$ to $(V-K)$ using
the color-color relations of \citet{bb88}, yielding $(V-K)_0=1.75$, and
then obtain equation (\ref{eqn:thetastar}) using
the color/surface-brightness relations of \citet{kervella04}. Combining equations (\ref{eqn:rho}) and (\ref{eqn:thetastar}) gives
$\theta_\e = \theta_*/\rho = 0.32\,$mas.  And combining this with the
definition $\theta_\e^2 = \kappa M\pi_\rel$, where $M$ is  the
lens  mass, $\pi_\rel$ is the source-lens relative parallax, and
$\kappa = 4G/c^2{\rm AU}\sim 8.1\,{\rm mas}\,M_\odot^{-1}$, together  with
the measured Einstein timescale, $t_\e=14.3\,\pm0.3$days, we obtain
\begin{equation}
M= 0.10 \pm 0.01 \,M_\odot\biggl({\pi_\rel\over 125\,\mu\rm as}\biggr)^{-1}
\label{eqn:mpirel}
\end{equation}
and 
\begin{equation}
\mu_\rel = {\theta_\e\over t_\e}=8.2 \pm 0.5 \,{\rm mas\,yr^{-1}}.
\label{eqn:propmot}
\end{equation}
The relatively high lens-source relative proper motion $\mu_\rel$ is
mildly suggestive of a foreground disk lens, but still quite consistent
with a bulge lens.
Since $\pi_\rel=125\,\mu\rm as$
corresponds to a lens distance $D_L=4\,\kpc$ (assuming source
distance $D_S=8\,\kpc$), equation (\ref{eqn:mpirel})
implies that if the lens did lie in the foreground, then it would be
a very low-mass star or a brown dwarf.

Assuming that the source lies at a Galactocentric distance
modulus 14.52, its dereddened color and magnitude imply
that $[(V-I)_0,M_I] = (0.81,3.07)$, making it a subgiant.

\section{Limb Darkening
\label{sec:ld}}

As illustrated in Figure~\ref{fig:lc}, the principal deviations
from a point-lens light curve occur at the limb of the star.
This prompts us to investigate the degree to which the planetary
solution is influenced by our treatment of limb darkening.  The
results that we report are based on a fit to the $H$-band surface
brightness profile of the form 
\begin{equation}
{S(\theta)\over S_0} = 1 
- \Gamma\biggl(1 - {3\over 2}\cos\vartheta\biggr)
- \Lambda\biggl(1 - {5\over 4}\cos^{1/2}\vartheta\biggr),
\label{eqn:ld}
\end{equation}
where $\Gamma$ and $\Lambda$ are the linear and square-root parameters,
respectively,  and where
$\vartheta$ is the angle between the normal to the stellar
surface and the line of sight, i.e., $\sin\theta = \vartheta/\theta_*$.
See \citet{an02} for the relation between $(\Gamma,\Lambda)$ and the
usual $(c,d)$ formalism.  

In deriving the reported results, we fix the $H$-band limb-darkening
parameters $(\Gamma,\Lambda) = (-0.15,0.69)$, corresponding to
$(c,d)=(-0.21,0.79)$ from \citet{claret00} for a star with
effective temperature $T_{\rm eff} = 5325\,$K and $\log g = 4.0$. 
These stellar parameters are suggested by comparison to 
Yale-Yonsei isochrones \citep{demarque04} for the dereddened color and 
absolute magnitude reported in \S~\ref{sec:finitesource}. We also 
perform fits in which $\Gamma$ and $\Lambda$ are allowed
to be completely free. From these fits, 
we find that our best-fit model has $(\Gamma,\Lambda) = (-0.64,1.47)$. 
$\Gamma$ and $\Lambda$ are highly correlated, so their individual values are not of 
interest, and the surface-brightness profiles generated by these two sets 
of $(\Gamma,\Lambda)$ are qualitatively similar. In the
present context, however, the key point is that when we fix the
limb-darkening parameters at the \citet{claret00} values, the contours in
Figure~\ref{fig:qgamma} remain essentially identical and the best fit
values change by much less than $1\,\sigma$.

Because of lower point-density and the aforementioned problems
with the $I$ data over the peak, we only attempt a linear
limb-darkening fit, i.e., we use equation~(\ref{eqn:ld}) with
$\Lambda\equiv 0$, and we adopt $\Gamma=0.47$ from \citet{claret00}.

\section{Blended Light
\label{sec:blend}}

In the crowded fields of the Galactic bulge, the photometered light
of a microlensing event rarely comes solely from the lensed source.  
Rather there is typically
additional light that is blended with the source but is not being
lensed.  This light can arise from unrelated stars that happen to be
projected close enough to the line of sight to be blended with the
source, or it can come from companions to the source, companions to
the lens, or the lens itself.  This last possibility is most
interesting because, if the lens flux can be isolated and measured,
it provides strong constraints on the lens properties, and in this
case would enable a complete solution of the lens mass and distance,
when combined with the measurement of $\theta_\e$ (e.g., \citealt{bennett07}).

To investigate the blended light, we begin by using the method
of \citet{gould02} to construct an image of the field with
the source (but not the blended light) removed, and compare this 
to a baseline image, which of course contains both the source
and the blended light.  In these images, the source/blend is immersed
in the wings of a bright star (roughly 3.7 mag brighter than the
source), which lies about $2''$ away.
On the baseline image, the source/blend is noticeable against this
background, but hardly distinct.  On the source-subtracted image,
the blend is not directly discernible.  

To make a quantitative estimate of the blend flux, we fit the
region in the immediate vicinity of the bright star to the form
$F = a_1 + a_2\times{\rm PSF}$, where ``PSF'' is the point-spread-function
determined from the DIA analysis.  We then subtract the best fit flux
profile from the image.  This leaves a clear residual at the
position of the source/blend in the baseline image, but just
noise in the source-subtracted image.  We add all the flux
in a $1.8''$ square centered on the lens, finding 564 ADU and
$-28$ ADU, respectively.  We conclude that the $I$-band flux
blend/source ratio is $f_b/f_s < 0.05$. Of course, even if we had 
detected blended light, it would be impossible to tell whether it 
was directly coincident with the source. If it were, this would imply that 
this light would be directly associated with the event, i.e.,
being either the lens itself
or a companion to the lens or the source.  Hence, this
measurement is an upper limit on the light from the lens in two
senses: no light is definitively measured, and if it were we do not
know that it came from the lens.
Combining this limit with equation (\ref{eqn:mpirel}), and
assuming the lens is a main-sequence star, it must then be less massive
than $M<0.75\,M_\odot$, and so must have relative parallax
$\pi_\rel>15\,\rm mas$.  This implies a lens-source separation
$D_S-D_L>1\,\kpc$, which certainly does not exclude bulge lenses.
Indeed, if the lens were a K dwarf in the Galactic bulge,
it would saturate this limit.

\section{Discussion
\label{sec:discuss}}

MOA-2007-BLG-400 is the first high-magnification microlensing event
for which the central caustic generated by a planetary companion to
the lens is completely enveloped by the source. As a comparison,
the planetary caustic of OGLE-2005-BLG-390 \citep{ob05390} is smaller 
than its clump-giant source star in angular size. When the planetary
caustics is covered by the source, the finite-source effects broaden
the ``classic'' \citet{gould92} planetary perturbation features \citep{gaudi97}.
By contrast, planet-induced deviations in MOA-2007-BLG-400 are mostly
obliterated, rather than being broadened, because the source crosses
the central caustic rather than the planetary caustic. We showed,
nevertheless, that the planetary character of the event can be
inferred directly from the light-curve features and that the standard
microlensing planetary parameters $(d,q)=(2.9,2.6\times 10^{-3})$ can
be measured with good precision, up to the standard close/wide
$d\leftrightarrow d^{-1}$ degeneracy.  We demonstrated that, in this
case, the close/wide degeneracy is quite severe, and the wide solution
is only preferred by $\Delta\chi^2 = 0.2$. This is unfortunate, since
the separations of the two solutions differ by a factor of $\sim 8.5$.
We argued that the severity of this degeneracy was primarily related
to the intrinsic parameters of the planet, rather than being
primarily a result of the large source size.

Although the mass ratio alone is of considerable interest for planet
formation theories, one would also like to be able to translate the
standard microlensing parameters to physical parameters, i.e., the
planet mass $m_p = qM$, and planet-star projected separation $r_\perp
= d\theta_\e D_L$.  Clearly this requires measuring the lens mass $M$ and distance $D_L$.
In this case, the pronounced finite source effects have already permitted 
a measurement of the Einstein radius $\theta_\e=0.32\,\mas$,
which gives a relation
between the mass and lens-source relative parallax (eq.~[\ref{eqn:mpirel}]).
This essentially yields a relation between the lens mass and distance,
since the source distance is close enough to the Galactic center that knowing $D_L$
is equivalent to knowing $\pi_{\rm rel}$.
Therefore, a complete solution could be determined by measuring either
$M$ or $D_L$, or some combination of the two.  

One way to obtain an independent relation between the lens mass and
distance is to measure the microlens parallax, $\pi_\e$.  
There are two potential ways of
measuring $\pi_\e$.  First, one can measure distortions in the light curve
arising from the acceleration of the Earth as it moves along its orbit.  Unfortunately, this is
out of the question in this case because the event is so short that
these distortions are immeasurably small.  Second, one can measure the
effects of terrestrial parallax, which gives rise to differences
between the light curves simultaneously observed from two or more
observatories separated by a significant fraction of the diameter of
the Earth.  Practically, measuring these differences requires a
high-magnification event, which would appear to make this event quite
promising.  Unfortunately, although we obtained simultaneous
observations from two observatories separated by several hundred
kilometers during the peak of the event, one of these datasets suffers 
from large systematic errors and an unknown time zero point, rendering it 
unusable for this purpose.

\begin{figure}
\epsscale{0.8}
\plotone{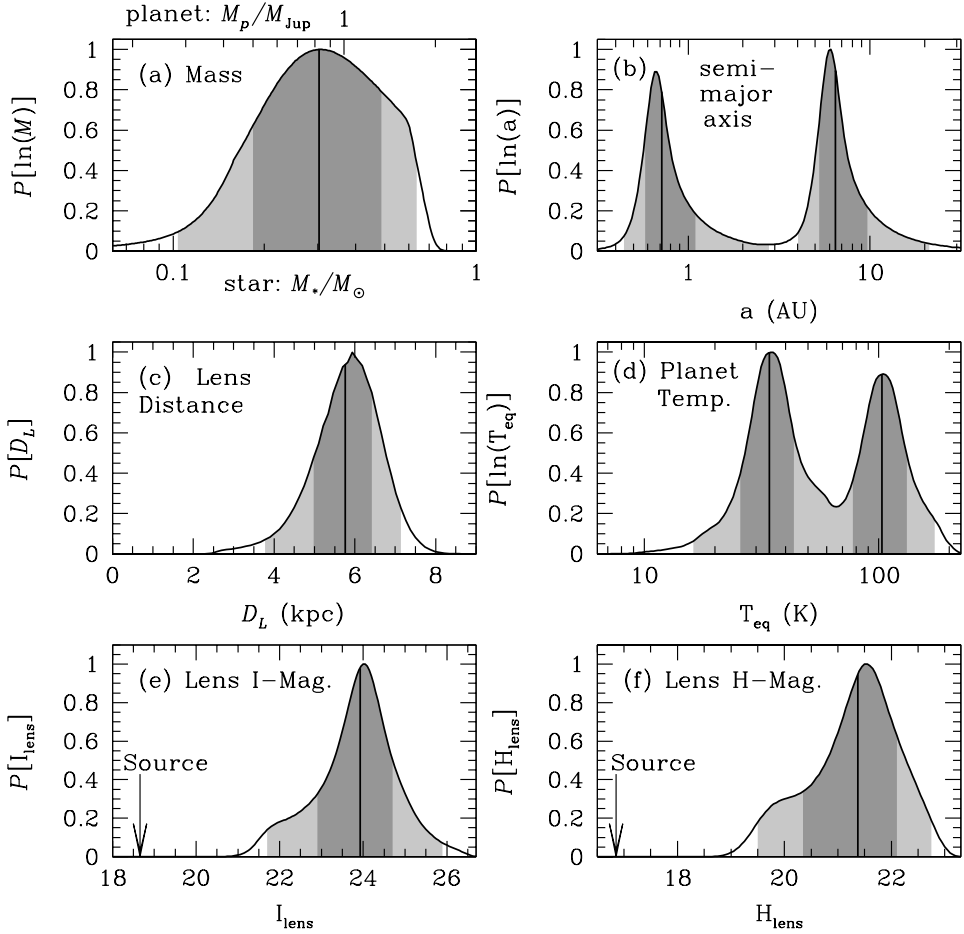}
\caption{Bayesian relative probability densities for the physical
properties of the planet MOA-2007-BLG-400Lb and its host star. (a)
Mass of the host star.  (b) Planet semimajor axis. (c) Distance to the
planet/star system. (d) Equilibrium temperature of the planet.  (e)
$I$-band magnitude of the host star. (f) $H$-band magnitude of the host star.
In panel (a), we also show the probability density for the planet mass,
which is essentially a rescaling of that of the host star, because
the mass ratio is measured so precisely $q=2.6\pm 0.4 \times 10^{-3}$.
In all panels, the solid vertical lines show the medians, and
the $68.3\%$ and $95.4\%$ confidence intervals are enclosed in the dark and
light shaded regions, respectively. In panel (b) and (d), the probability
distributions for wide and close degenerate solutions are computed separately, 
and then the wide solution is weighted by $exp(-\Delta\chi^2/2)$, where 
$\Delta\chi^2 = 0.2$ is the difference between them.
These distributions are derived assuming priors obtained from standard
models of the mass, velocity, and density distributions of stars in
the Galactic bulge and disk, and include constraints from the
measurements of lens angular Einstein radius $\theta_\e$ and the
timescale of the event $t_\e$, as well as limits on the 
$I$-band magnitude of the lens.  In practice, only the
measurements of $\theta_\e$ and $t_\e$ provide 
interesting constraints on these distributions.}
\label{fig:prop}
\end{figure}

The only available alternative for breaking the degeneracy
between the lens mass and distance would be to measure the
lens flux, either under the glare of the source or, at a later
date, to separately resolve it after it has moved away from the 
line of sight to the source \citep{alcock01,koz07}.  
Panels (e) and (f) of Figure~\ref{fig:prop} show the Bayesian estimates of 
the lens brightness in $I$-band and $H$-band, respectively. If the lens flux is at least 2\% of the source flux, then the former
kind of measurement could be obtained from a single epoch {\it Hubble Space
Telescope} observation, provided it were carried out in the
reasonably near future.  At roughly 99\% probability, the blended light 
would be either perfectly aligned with the source (and so associated
with the event) or well separated from it.
{\it HST} images can be photometrically
aligned to the ground-based images using comparison stars with 
an accuracy of better than 1\%.  Hence, photometry of the source+blend
would detect the blend, unless it were at least 4 mag fainter than the
source.  In principle, the blend could be a companion to either the
source or lens.  Various arguments can be used to constrain 
either of those scenarios.  We do not explore those here, but see
\citet{dong08}.
If the lens is not detectable by current epoch {\it HST} observations
(or no {\it HST} observations are taken), then it will be detectable
by ground-based AO $H$-band observations in about 5 years.  This
is because the lens-source relative proper motion is measured to be
$\mu_\rel = 8\,\mas\,\rm yr^{-1}$, and the diffraction limit at $H$ band
on a 10m telescope is roughly 35 mas.  If the lens proves to be
extremely faint, then a wider separation (and hence a few years more
time baseline) would be required.  

In the absence of additional observational constraints, we must
rely on a Bayesian analysis to estimate the properties of the host
star and planet, which incorporates priors on the distribution
of lens masses, distances, and velocities \citep{dominik06,ob04343}. 
This is a standard procedure, which we only briefly summarize here. 
We adopt a \citet{han95} model for the Galactic bar,
a double-exponential disk with a scale
height of 325 pc, and a scale length of 3.5 kpc, as well as other
Galactic model parameters as described in \citet{bennett02}.
We incorporate constraints from our 
measurement of the lens angular Einstein radius $\theta_\e$ and the
event timescale, as well as limits on the microlens parallax and
$I$-band magnitude of the lens.  In practice, only the measurements of $\theta_\e$ and $t_{\e}$ provide interesting constraints on 
these distributions. In addition, we include the small penalty on the close 
solution, $\exp(-\Delta \chi^2/2)$, where the wide solution is favored by 
$\Delta\chi^2 = 0.2$. For the estimates
of the planet semimajor axis, we assume circular orbits and that the orbital
phases and cos(inclinations) are randomly distributed.  

The resulting probability densities for the physical
properties of the host star, as well as selected properties
of the planet, are shown in Figure~\ref{fig:prop}.  The Bayesian
analysis suggests a host star of mass $M=0.30_{-0.12}^{+0.19} M_\odot$
at distance of $D_L=5.8_{-0.8}^{+0.6}~{\rm kpc}$.  In other words,
given the available constraints, the host is most likely an M-dwarf, probably 
in the foreground Galactic bulge.  Given that the planet/star
mass ratio is measured quite precisely, the probability distribution
for the planet mass is essentially just a rescaled version of the
probability distribution for the host star mass. We find
$m_p=0.82_{-0.33}^{+0.52}~M_{\rm Jup}$.  The close/wide degeneracy
is apparent in the probability distribution for the semimajor axis $a$.
We estimate
$a_{\rm close}= 0.72_{-0.16}^{+0.38}~{\rm AU}$ for the close solution, and 
$a_{\rm wide}=6.5_{-1.2}^{+3.2}~{\rm AU}$ for the wide solution.  The
equilibrium temperatures for these orbits are 
$T_{\rm eq., close}=103_{-26}^{+28}~{\rm K}$ and 
$T_{\rm eq., wide}=34\pm{9}~{\rm K}$
for the close and wide solutions, respectively.  

Thus our Bayesian analysis suggests that this system is mostly likely
a bulge mid-M-dwarf, with a Jovian-mass planetary companion.  The
semimajor axis of the planetary companion is poorly constrained
primarily because of the close/wide degeneracy, but the implied
equilibrium temperatures are cooler than the condensation temperature of water. 
Specifically we find that $T_{\rm eq}
\la 173~{\rm K}$ at $2\,\sigma$ level.  Alternatively, if we assume
the snow line is given by $a_{\rm snow}=2.7~{\rm AU}(M/M_\odot)$, we find
for this system a snow line distance of $\sim 0.81~{\rm AU}$,
very close to the inferred semimajor axis of the close
solution.  Thus this planet is quite likely to be located close to or 
beyond the snow line of the system.

Although we cannot distinguish between the close and wide solutions
for the planet separation, theoretical prejudice in the context of the
core-accretion scenario would suggest that a gas-giant planet would be
more likely to form just outside the snow line, thus preferring the
close solution.  However, we have essentially no observational constraints on the
frequency and distribution of Jupiter-mass planets at the separations
implied by the wide solution ($\sim 5.3-9.7~{\rm AU}$), for such low-mass
primaries.  Unfortunately, the prospects for empirically resolving the close/wide
degeneracy in the future are poor. The only possible method of doing
this would be to measure the radial velocity signature of the planet.
Given the faintness of the host star (see \S\ref{sec:blend} and Fig.\
\ref{fig:prop}), this will likely be impossible with current or
near-future technology.

The mere existence of a gas-giant planet orbiting a mid-M-dwarf is
largely unexpected in the core-accretion scenario, as formation of
such planets is thought to be inhibited in such low-mass primaries
\citep{laughlin04}.  Observationally, however, although the frequency
of Jovian companions to M-dwarfs with $a\la 3~{\rm AU}$ does appear to
be smaller than the corresponding frequency of such companions to FGK
dwarfs \citep{endl06,johnson07,cumming08}, several Jovian-mass
companions to M dwarfs are known (see \citealt{dong08} for a
discussion), so this system would not be unprecedented.  Furthermore,
it must be kept in mind that the estimates of stellar (and so planet)
mass depend on the validity of the priors, and even in this context have
considerable uncertainties. 

Most of the ambiguities in the interpretation of this event would
be removed with a measurement of the host star mass and distance,
which could be obtained by combining our measurement of $\theta_\e$
with a measurement of the lens light as outlined above.  The Bayesian 
analysis informs the likelihood of success of such an endeavor. This analysis
suggests that, if the host is a main-sequence star, its magnitude will be 
$I_L=23.9_{-1.0}^{+0.8}$ and $H_L=21.4_{-1.0}^{+0.7}$, which corresponds
to $0.6\%$ and $1.7\%$ of the source flux, respectively. 
If initial efforts to
detect the lens fail, more aggressive observations would certainly
be warranted: microlensing is the most sensitive method for detecting
planets around very low-mass stars simply because it is the only
method that does not rely on light from the host (or the planet itself)
to detect the planet.
And given equation~(\ref{eqn:mpirel}), even an M dwarf at the very
bottom of the main sequence $M=0.08\,M_\odot$, would lie at $D_L=3.5\,\kpc$
and so would be $H\sim 24$.

\acknowledgments

We thank Shude Mao, David Heyrovsky and Pascal Fouque for useful comments 
during the completion of this work. 
S.D. and A.G. were supported in part by grant AST-0757888 from the NSF. S.D., 
A.G., D.D. and R.P. acknowledge support by NASA grant NNG04GL51G. 
The MOA group acknowledge Ministry of Education, Culture, Sports,
Science and Technology of Japan (MEXT and JSPS) no.18253002 and 20340052
for support. IAB,JBH,PMK,DJS,WS,PCMY acknowledge Marsden Fund of NZ.
D.P.B. was supported by grants AST-0708890 from the NSF and NNX07AL71G 
from NASA. LS is supported by NZ Foundation of Science, Research, and 
Technology. CH is supported by SRC Korea Science \& Engineering Foundation. 
B-GP acknowledge the support by Korea Astronomy \& Space Science Institute.
The OGLE project is partially supported by the Polish MNiSW grant
N20303032/4275. This work was supported in part by an allocation of 
computing time from the Ohio Supercomputer Center.

\appendix
\section{Failure of Elliptical-Source Models}
\label{sec:append_ell}

\begin{figure}
\epsscale{0.9}
\plotone{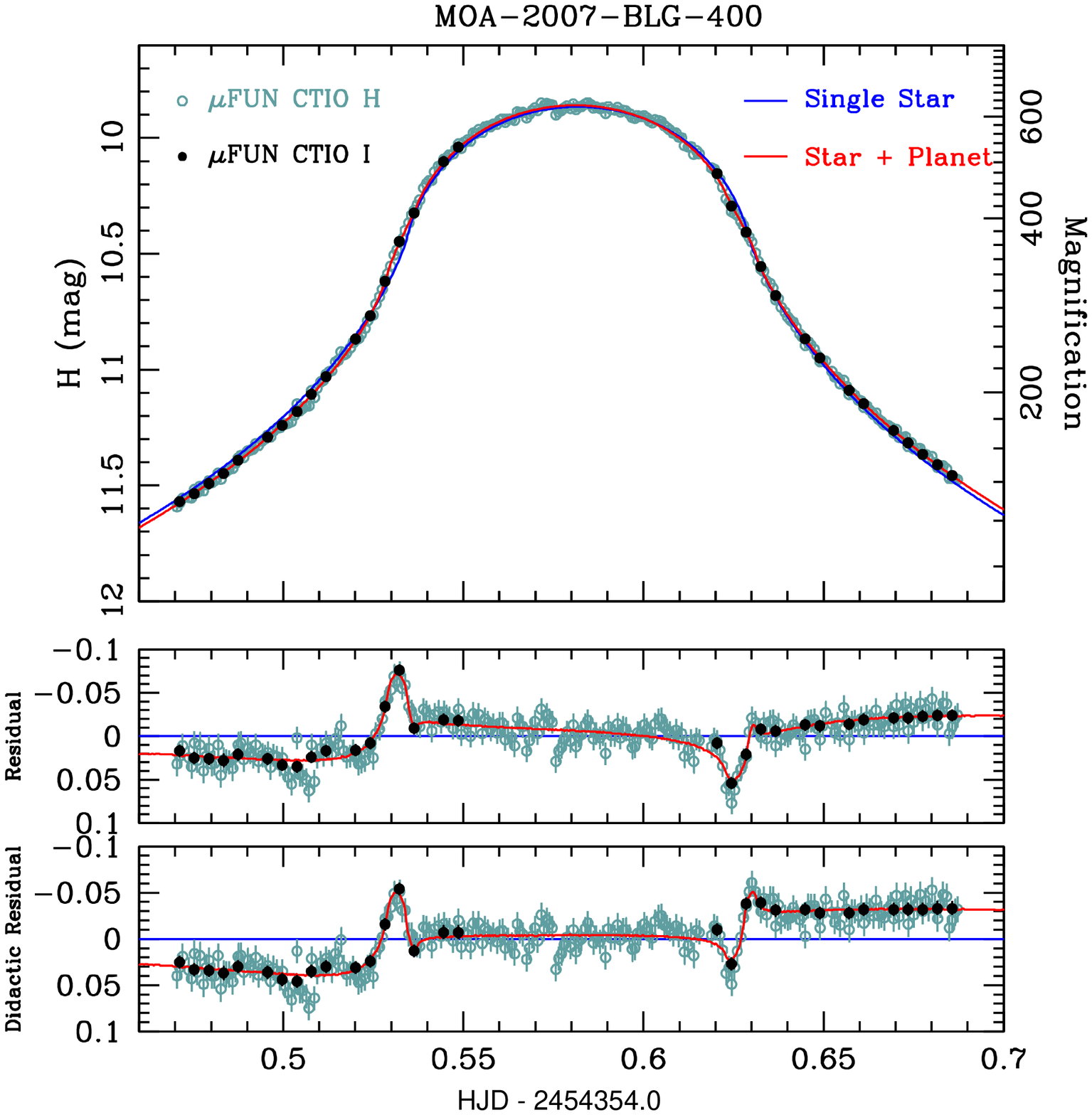}
\caption{Top: Lightcurve of MOA-2007-BLG-400 with data from $\mu$FUN CTIO
(Chile) simultaneously taken in $H$ ({\it cyan}),  
$I$ (DIA, {\it black}). Models are shown
for a point lens ({\it blue}) and planet-star system ({\it red}).
There are 5 50-second $H$ exposures for each 300-second
$I$ (or $V$ -- not shown) exposure in 6 minutes cycles. Some $I$-band data at 
the peak suffer from saturation, and those points are therefore 
removed from the analysis (see text). 
Middle: Residuals for best-fit point-lens model and its difference with the 
planetary model.  Note that in the top panel,
the $H$ data are shown as observed, while the $I$ data are aligned.
Normally, such alignment is straightforward because microlensing
of point sources is achromatic.  However, here there is significant
chromaticity due to different limb-darkening.  The $I$-band points
in the top panel are actually the residuals to the $I$-band limb-darkened
model (middle panel), added to the $H$-band model curve (top panel). 
Bottom: Residuals from a point-lens model with the same parameters as the 
planetary model, which can be directly compared to the ``magnification 
map'' in Fig.~\ref{fig:caustic}. These ``didactic residuals'' are naturally more pronounced than those from the best-fit point lens.
}
\label{fig:lc}
\end{figure}

\begin{figure}
\epsscale{0.9}
\plotone{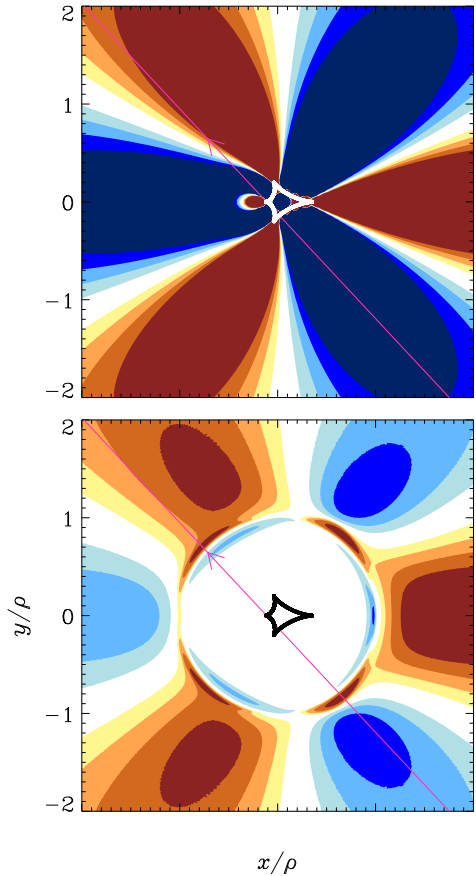}
\caption{
Magnification differences between of best-fit planetary model
[$(q,d)=(0.0026,2.9)$ and $(q,d)=(0.0026,0.34)$ being nearly
identical] and single-lens models, in units of the measured
source size, $\rho = 0.0033$ Einstein radii. 
Contours show 1\%, 2\%, 3\%, and 4\%, deviations in the positive ({\it brown})
and negative ({\it blue}) directions.
Top panel: Single-lens geometry ($t_0,u_0,t_\e$) is taken to be the
same as in the planetary model, with no finite source effects.
Caustic (contour of infinite magnification) is shown in white.
The deviations are very pronounced. Bottom Panel: Same as top
panel, but including finite-source effects, which now explain the
main features of the light curve.
The trajectory begins with a negative deviation, then
hits a narrow ``brown ridge'' causing the spike seen in the bottom
panel of Fig.~\ref{fig:lc},
as the edge of the source first hits the caustic.  Then there are
essentially no deviations ({\it white}) while the source covers the
caustic. The caustic exit induces a narrow ``blue ridge'' corresponding
to the negative-deviation spike seen in Fig~\ref{fig:lc}.  Finally,
the source runs along the long ``brown ridge'' corresponding to the
prolonged post-peak mild excess seen in Fig~\ref{fig:lc}.
}
\label{fig:caustic}
\end{figure}

\begin{figure}
\plotone{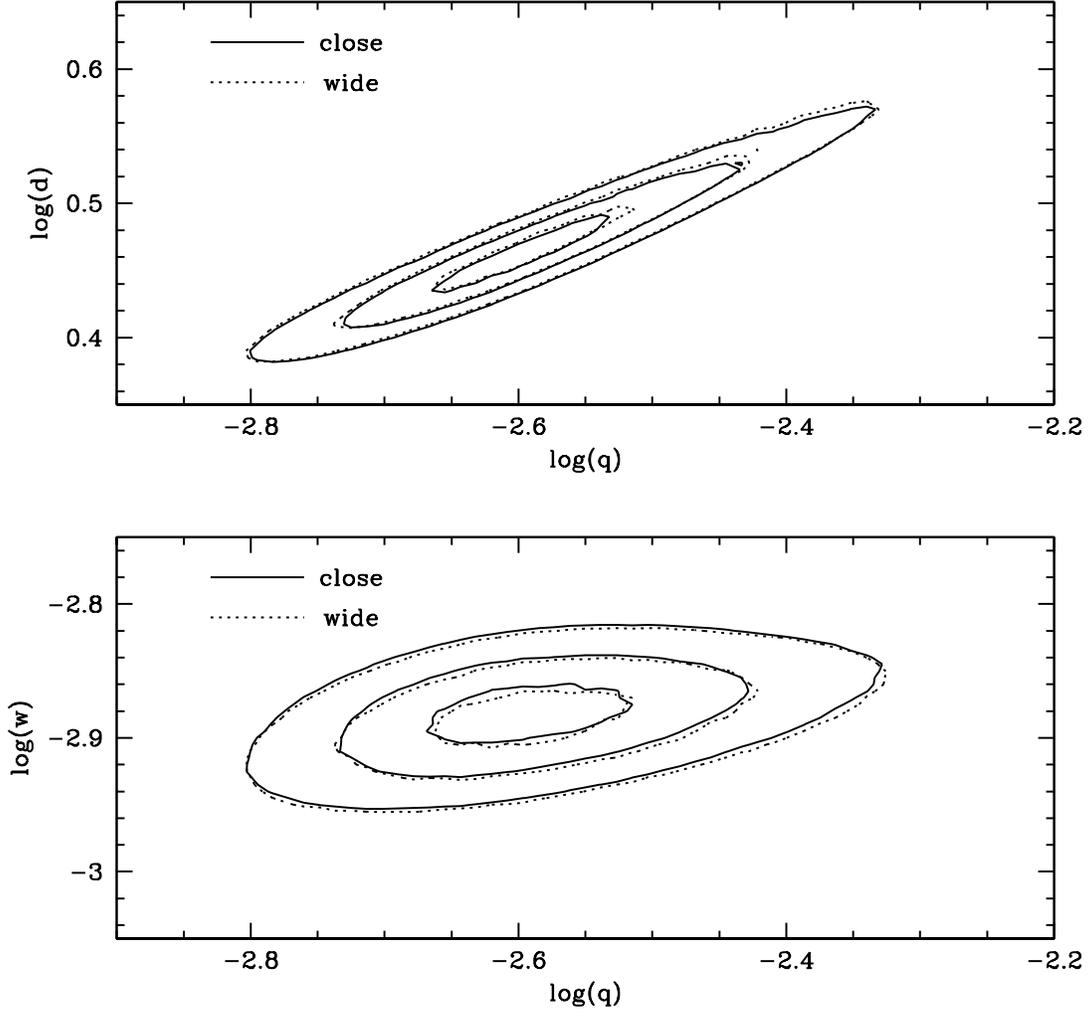}
\caption{Contours of $\Delta\chi^2=1,\,4,\,9$ relative to the
minimum as a function of planet-star mass ratio $q$ and projected 
planet-star separation $d$ (top), as well as 
``short caustic diameter'' (see Fig.~\ref{fig:caustic}) $w$ (bottom).
$w$ (in units of $\theta_\e$) is a function of $q$ and $d$
(see text).  The solution shown here corresponds to 
$q=2.6\pm 0.4\times 10^{-3}$ and $d=2.9\pm 0.2$ (or $d=0.34\pm 0.02$).
These values of $d$ correspond to physical separations and equilibrium
temperatures of
$\sim 0.6-1.1~{\rm AU}$, $\sim 103~{\rm K}$ and 
$\sim 5.3-9.7~{\rm AU}$, $\sim 34~{\rm K}$ for the close and wide solutions,
respectively.}
\label{fig:qgamma}
\end{figure}

Because MOA-2007-BLG-400 is the first microlensing event with a completely
buried central caustic, it is important to rule out other potential causes of
the deviations seen in the light curve (apart from a planetary companion
to the lens).  The principal features of these deviations are the
twin ``spikes'' in the residuals, which are approximately centered
on the times when the lens enters and exits the source.  In the model,
these crossings occur at about HJD' $= 4354.53$ and HJD' $=4354.63$,
i.e., very close to the spikes in Figure~\ref{fig:lc}.  In principle,
one might be able to induce such spikes by displacing the model
source crossing times from the true times.  The only real way
to achieve this (while still optimizing the overall fit parameters)
would be if the source were actually elliptical, but were modeled
as a circle (which, of course, is the norm).

One argument against this hypothesis is the similarity of
the $I$ and $H$ residuals (\S~\ref{sec:model}). If the source
were an ellipsoidal variable, then one would expect color
gradients due to ``gravity darkening''.

Nevertheless, we carried out two types of investigation of this possibility.
First, we modeled the light curve as an elliptical source magnified
by a point (non-binary) lens.  In addition to the linear flux parameters
(source flux plus blended flux for each observatory) there are
6 model parameters, the three standard point-lens parameters ($t_0,u_0,t_\e$),
plus the source semimajor and semiminor axes ($\rho_a$, $\rho_b$) and the
angle of the source trajectory relative to the source major axis, $\alpha$.
We find that the elliptical source reduces $\chi^2$ by about 200,
but it does not remove the ``spikes'' from the residuals, which was
the primary motivation for introducing it.  Instead, essentially
all of the $\chi^2$ improvement comes from eliminating the
asymmetries from the rest of the light curve.  Recall, however,
that the planetary model removes both these asymmetries and
the ``spikes''.  Moreover, the best-fit axis ratio is quite 
extreme, $\rho_b/\rho_a = 0.7$, which would produce very noticeable
ellipsoidal variations unless the binary were being viewed
pole on.

Next we looked for sinusoidal variations in the baseline light curve.
The individual OGLE errorbars at baseline are smaller than for MOA, and 
since ellipsoidal variations are strictly periodic, the longer OGLE 
baseline (about $T=2000$ days versus $T=800$ days for MOA) does a better 
job of isolating this signal from various possible systematics. Therefore, 
for this purpose, the OGLE data are more suitable than MOA.
The OGLE data are essentially all baseline (only two
magnified points out of 452).  Their periodogram shows several spikes
at the 0.01 mag level, and a maximum $\Delta\chi^2=20$,
which are consistent with noise.   The width of the spikes is
extremely narrow, consistent with the theoretical expectation
for uniformly sampled data of
$\sigma(P)/P^2\sim 
\sqrt{24/\Delta\chi^2}/(2\pi T)\sim 10^{-4}\,\rm day^{-1}$,
indicating that the data set is behaving normally.

In brief, our investigation finds no convincing evidence for ellipticity
of the source, certainly not for the several tens of percent deviation
from circular that would be needed to significantly ameliorate 
the deviations seen near peak in the light curve.
Moreover, even arbitrary source ellipticities cannot reproduce the light curve's
most striking features: the two ``spikes'' in the residuals that occur
when the lens crosses the source boundary.

{\bf
\begin{deluxetable}{l r r r r r r r r r r r c}
\tablecaption{\label{tab:models} Best-fit Planetary Models}
\tablewidth{0pt} \tablehead{
\colhead{Model\tablenotemark{1}} &
\colhead{$t_0-t_{\rm ref}$\tablenotemark{2}} &
\colhead{$u_0$} &
\colhead{$t_\e$} &
\colhead{$d$} &
\colhead{$q$} &
\colhead{$\alpha$\tablenotemark{3}} & 
\colhead{$\rho$}  
\\
\colhead{} &
\colhead{day} &
\colhead{} &
\colhead{day} &
\colhead{} &
\colhead{} &
\colhead{deg} &
\colhead{} &
}
\startdata

Close & 0.08107 & 0.00025 & 14.41 & 0.34 & 0.0026 &  227.06 & 0.00326\\
Wide  & 0.08106 & 0.00027 & 14.33 & 2.87 & 0.0025 &  226.99 & 0.00329\\
\enddata
\tablenotetext{1}{The wide solution is preferred over best-fit single-lens
model by $\Delta{\chi^2} = 1070.04$ and the close solution by $\Delta{\chi^2} = 1069.84$}
\tablenotetext{2}{$t_{\rm ref}=$ HJD $2454354.5$}
\tablenotetext{3}{The geometry of the source trajectory is illustrated in Figure \ref{fig:caustic}, in which the planet is to the right of the lens star. $t_0$,
$u_0$, and $\alpha$ are defined with respect to the ``center of magnification'',
which is the center of mass of the star/planet system for the close model and $q/(1+q)/d$ away from the position of the lens star toward the direction 
of the planet for the wide model.
}
\end{deluxetable}
}

\begin{thebibliography}{99}

\bibitem[Albrow et al.(2002)]{albrow02} Albrow, M.~D., et al.\ 
2002, \apj, 572, 1031 

\bibitem[Alcock et al.(2001)]{alcock01}Alcock, C., et al. 2001, Nature, 414, 617

\bibitem[An et al.(2002)]{an02} An, J. 2002, \apj, 572, 521

\bibitem[Beaulieu et al.(2006)]{ob05390} Beaulieu, J.-P. et al. 2005,
Nature, 439, 437

\bibitem[Bennett 
\& Rhie(1996)]{bennett96} Bennett, D.~P., \& Rhie, S.~H.\ 1996, \apj, 472, 660 

\bibitem[Bennett 
\& Rhie(2002)]{bennett02} Bennett, D.~P., \& Rhie, S.~H.\ 2002, \apj, 574, 985 

\bibitem[Bennett et al.(2007)]{bennett07} Bennett, D.~P., 
Anderson, J., \& Gaudi, B.~S.\ 2007, \apj, 660, 781 

\bibitem[Bennett et al.(2008)]{mb07192} Bennett, D.P. et al.,\ 2008, \apj, 684, 663

\bibitem[Bessell \& Brett(1988)]{bb88}
Bessell, M. S., \& Brett, J. M.\ 1988, \pasp, 100, 1134

\bibitem[Bond et al.(2001)]{bond01} Bond, I.~A., et al.\ 2001, 
\mnras, 327, 868 

\bibitem[Bond et al.(2002)]{bond02}Bond, I. A., et al.\ 2002, \mnras, 331, L19

\bibitem[Bond et al.(2004)]{ob03235}
Bond, I.A., et al.\ 2004, \apj, 606, L155

\bibitem[Burrows et al.(1993)]{burrows93} Burrows, A., Hubbard, 
W.~B., Saumon, D., \& Lunine, J.~I.\ 1993, \apj, 406, 158 

\bibitem[Burrows et al.(1997)]{burrows97} Burrows, A., et al.\ 
1997, \apj, 491, 856 

\bibitem[Chang-Refsdal(1979)]{cr1} 
Chang, K.\ \& Refsdal, S.\ 1979, Nature, 282, 561

\bibitem[Chung et al.(2005)]{chung05} Chung, S.-J. et al, 2005, \apj, 630, 535

\bibitem[Claret(2000)]{claret00}Claret A. 2000 \aap, 363, 1081

\bibitem[Cumming et al.(2008)]{cumming08} Cumming, A., Butler, 
R.~P., Marcy, G.~W., Vogt, S.~S., Wright, J.~T., 
\& Fischer, D.~A.\ 2008, \pasp, 120, 531 

\bibitem[Demarque et al.(2004)]{demarque04} Demarque, P., Woo, 
J.-H., Kim, Y.-C., \& Yi, S.~K.\ 2004, \apjs, 155, 667 

\bibitem[Dominik(1999)]{dominik99} 
Dominik, M.\ 1999, \aap, 349, 108

\bibitem[Dominik(2006)]{dominik06} Dominik, M.\ 2006, \mnras, 
367, 669 

\bibitem[Dong et al.(2006)]{ob04343} Dong, S., et al. 2006, \apj, 642, 842

\bibitem[Dong et al.(2008)]{dong08} Dong, S., et al. 2008, \apj, submitted
(arXiv:0804.1354)

\bibitem[Endl et al.(2006)]{endl06} Endl, M., Cochran, W.~D., 
K{\"u}rster, M., Paulson, D.~B., Wittenmyer, R.~A., MacQueen, P.~J., 
\& Tull, R.~G.\ 2006, \apj, 649, 436 

\bibitem[Gaudi 
\& Gould(1997)]{gaudi97} Gaudi, B.~S., \& Gould, A.\ 1997, \apj, 486, 85 

\bibitem[Gaudi et al.(2008)]{ob06109} 
Gaudi, B. S., et al.\ 2008, Science, 315, 927

\bibitem[Gould(2008)]{gould08} 
Gould, A.\ 2008, \apj, 681, 1593

\bibitem[Gould \& An(2002)]{gould02} 
Gould, A.\ \& An, J.H. 2002, \apj, 565, 1381

\bibitem[Gould 
\& Loeb(1992)]{gould92} Gould, A., \& Loeb, A.\ 1992, \apj, 396, 104 

\bibitem[Gould et al.(2006)]{ob05169} Gould, A., et al. 2004, \apj, 644, L37

\bibitem[Griest \& Safizadeh(1998)]{griestsafi}
Griest, K.\ \& Safizadeh, N.\ 1998, \apj, 500, 37

\bibitem[Han(2007)]{han07} Han, C.\ 2007, \apj, 661, 1202 

\bibitem[Han \& Gaudi(2008)]{han08} Han, C., \& Gaudi, B.~S.\ 2008, ArXiv e-prints, 805, arXiv:0805.1103 

\bibitem[Han 
\& Gould(1995)]{han95} Han, C., \& Gould, A.\ 1995, \apj, 447, 53

\bibitem[Ida 
\& Lin(2005)]{ida05} Ida, S., \& Lin, D.~N.~C.\ 2005, \apj, 626, 1045 

\bibitem[Johnson et al.(2007)]{johnson07} Johnson, J.~A., Butler, 
R.~P., Marcy, G.~W., Fischer, D.~A., Vogt, S.~S., Wright, J.~T., 
\& Peek, K.~M.~G.\ 2007, \apj, 670, 833 

\bibitem[Kennedy et al.(2006)]{kennedy06} Kennedy, G.~M., Kenyon, 
S.~J., \& Bromley, B.~C.\ 2006, \apjl, 650, L139 

\bibitem[Kennedy 
\& Kenyon(2008)]{kennedy08} Kennedy, G.~M., \& Kenyon, S.~J.\ 2008, \apj, 673, 502 

\bibitem[Kervella et al.(2004)]{kervella04} Kervella, P., 
Th{\'e}venin, F., Di Folco, E., \& S{\'e}gransan, D.\ 2004, \aap, 426, 297 

\bibitem[Koz\l owski et al.(2007)]{koz07} Koz\l owski, S., Wo\'zniak, P.R.,
Mao, S., \& Wood, A. 2007, \apj, 671, 420

\bibitem[Laughlin et al.(2004)]{laughlin04} Laughlin, G., 
Bodenheimer, P., \& Adams, F.~C.\ 2004, \apjl, 612, L73 

\bibitem[Lecar et al.(2006)]{lecar06} Lecar, M., Podolak, M., 
Sasselov, D., \& Chiang, E.\ 2006, \apj, 640, 1115 

\bibitem[Liebes(1964)]{liebes} Liebes, S.\ 1964, Physical 
Review , 133, 835 

\bibitem[Lin et al.(1996)]{lin96} Lin, D.~N.~C., Bodenheimer, 
P., \& Richardson, D.~C.\ 1996, \nat, 380, 606 
 
\bibitem[Lissauer(1987)]{lissauer87} Lissauer, J.~J.\ 1987, 
Icarus, 69, 249 

\bibitem[Mao \& Paczynski(1991)]{mp91} Mao, S., \& Paczynski, B.\ 1991, \apjl, 374, L37 


\bibitem[Park et al.(2006)]{park06} Park, B.-G., Jeon, Y.-B., 
Lee, C.-U., \& Han, C.\ 2006, \apj, 643, 1233

\bibitem[Pejcha \& Heyrovsky(2007)]{pejcha07} 
Pejcha, O., \& Heyrovsky, D.\ 2007, ArXiv e-prints, 712, arXiv:0712.2217

\bibitem[Pollack et al.(1996)]{pollack96} Pollack, J.~B., 
Hubickyj, O., Bodenheimer, P., Lissauer, J.~J., Podolak, M., 
\& Greenzweig, Y.\ 1996, Icarus, 124, 62 

\bibitem[Rasio 
\& Ford(1996)]{rasio96} Rasio, F.~A., \& Ford, E.~B.\ 1996, Science, 274, 954 

\bibitem[Rattenbury et al.(2002)]{rattenbury02}
Rattenbury, N. J., Bond, I. A., Skuljan, J., \& Yock, P. C. M. 2002, \mnras,
335, 159

\bibitem[Schechter et al.(1993)]{dophot}
Schechter, P.L., Mateo, M. \& Saha, A.  1993, \pasp, 105 1342

\bibitem[Udalski et al.(2005)]{ob05071}
Udalski, A., et al. 2005, \apj, 628, L109

\bibitem[Ward(1997)]{ward97} Ward, W.~R.\ 1997, Icarus, 126, 
261 

\bibitem[Wo\'zniak(2000)]{wozniak}
Wo\'zniak, P.R. 2000, Acta Astron., 50, 421

\bibitem[Yoo et al.(2004)]{ob03262} Yoo, J., et al.\ 2004, \apj, 603, 139

\end{thebibliography}
\end{document}